\documentclass[a4paper,11pt,twoside,onecolumn]{article}
\usepackage[utf8]{inputenc} 
\usepackage{color}
\usepackage{graphicx}
\usepackage{amsmath}
\usepackage{amssymb}
\usepackage[bottom=3cm,top=3cm,left=3cm,right=3cm]{geometry}
\usepackage{float}

\newcommand{\be}{\begin{equation}}
\newcommand{\ee}{\end{equation}}
\newcommand{\ben}{\begin{equation*}}
\newcommand{\een}{\end{equation*}}

\usepackage{fancyvrb}
\usepackage{xfrac}

\newcommand{\vb}[1]{\mathbf{#1}} 
\newcommand{\euler}[1]{\mathrm{e}^{#1}} 

\begin{document}

\markboth{Blanc, Gonzalez, Lavia}
{Modelling High-Frequency Backscattering from a Mesh of Curved Surfaces Using Kirchhoff Approximation}


\title{Modelling High-Frequency Backscattering from a Mesh of Curved Surfaces Using Kirchhoff Approximation}
\author{Edmundo Lavia, Juan D. Gonzalez, Silvia Blanc}

\maketitle

\begin{abstract}
The Kirchhoff approximation (K-A) to calculate the acoustic backscattering of a complex structure can be evaluated 
using a discretized version of its surface (i.e. a \textit{mesh}). From the computational viewpoint, the most handy 
approach is the one based on flat facets. However, in the high frequency range, where the K-A provides good agreement 
and is therefore applicable, it requires a mesh with such a large number of facets that it turns impractical. To avoid 
these difficulties a mesh of curved triangles can be used to model the scatterer's complex structure. Previous 
computational implementations reported in the literature did not accomplish satisfactory results for high frequency.
In this work we propose a numerical model based upon an iterative integration using Gauss-Legendre rules.
The model was validated against exact solutions and led us to achieve adequate results in the high-frequency range. 
\end{abstract}

\section{Introduction}
\label{intro}


The Kirchhoff approximation (K-A) is probably the {\it quintessential} tool to calculate approximate high-frequency 
scattering from impenetrable bodies since it provides not only a computationally easy implementation but even a faster 
one.
The K-A allows for obtaining  the far-field backscattering amplitude function  $f_\infty$ through an integration over 
the insonified scatterer surface $S_i$ \cite{MedwinClay}, 
\begin{equation}  
	f_\infty = \frac{ik}{2\pi}\int_{S_{i}} \:\mathrm{e}^{2ik\hat{k}\cdot\vb{x}} \: 
	\hat{k}\cdot\hat{n}(\vb{x}) \: dS(\vb{x}), 
	\label{kirchoff1} 
\end{equation} 
where $ k $ is the wavenumber of the incident field, $\hat{k}$ its incidence direction and $\hat{n}$ the exterior normal 
to the scatterer surface.
When the object is convex, this approximation can be used to calculate the scattering provided  the frequency is higher 
enough ($k a \gg 1$, being $a$ a characteristic length of the scatterer).
For non convex bodies, multiple scattering must be taken into account in order to allow that the K-A still works. 
Furthermore, in that case shadowing algorithms are needed to determine which region of the object surface is insonified 
by the incoming wave and its successive reflections. 

Several methods have been developed over the years to model more accurately the high-frequency scattering by 
supplementing the K-A with the accounting of edge diffraction \cite{Fawcett}, material properties through reflection 
coefficients \cite{Lee} and multiple scattering contributions. These issues (non convexities, shadowing and diffraction 
corrections) are out of the scope of this work and consequently, only the convex case and the bare K-A will be 
considered here.

When the surface of a scatterer is easily described, as in the case of spheres, cylinders or parallelepipeds, the 
integration can be carried out by analytical methods and sometimes it has a closed-form \cite{Gaunard}. On the other 
hand, when the scatterer has a complex geometry (e.g. fish, phytoplankton or submarine hulls), an usual approach 
consists of approximating its surface by a mesh composed of simpler elements as triangular or quadrilateral facets 
(planar or curved).
Then, using the linearity of the integral, the K-A reduces to the sum of the individual integrals over each insonified 
element.

There are many works where the approach based on planar triangular facets is applied. For example, George 
\cite{George-PlaneFacets} has used analytical approximations that may introduce certain lack of precision, whereas 
Pignier \textit{et al.}\cite{Pignier} have used quadrature rules but they led to computationally expensive expressions.
Other authors explicitly use the fact that {the integral} in Eq. \eqref{kirchoff1} has an exact solution based 
on elementary functions when the {integration surface (the facet)} is a planar triangle. Therefore, the 
computational cost is reduced to a minimum when {the integrations implied by Eq. \eqref{kirchoff1} are carried 
out in this fasion}\cite{Sammelman}$^{,}$ \cite{Wendelboe}$^{,}$  \cite{Abawi-Kirchhoff}. 

However, since flat facets cannot exactly represent any curvature, the description of a curved surface with this 
kind of mesh is only a approximation, although it is a very good one if the number of facets is big enough. 
It can be shown that higher the frequency bigger the required mesh size.
The advantage of the simple integration provided by the model of planar triangles is spoiled in the high frequency 
regime because of  the required mesh size, which makes impractical the computational implementation of Eq. 
\eqref{kirchoff1}. For this reason, difficulties arise even for  $ka > 20$, as stated in the literature 
\cite{Abawi-Kirchhoff}.
Other researchers provide numerical results that reach $ ka \approx 42 $  and express that the method potentiality 
reaches up to $ ka \approx 200$, although results for this case are not shown \cite{Wendelboe}.

Several authors have pointed out the issue that arises when attempting to represent a curved surface with planar facets 
(the so called {\it curvature} problem). With the aim of overcoming it, curvilinear facets may be employed, but in this 
case there are no longer analytical solutions thus quadrature rules should be used.
Some authors who follow this approach have considered seven-point Gaussian quadrature in 2D \cite{penia,footeFrancis}, 
nevertheless this kind of quadrature rules are not appropriate at high frequencies because the oscillatory behavior of 
the integrand in Eq \eqref{kirchoff1} causes numerical errors.
Therefore, this method is unable to provide high $ka$ values. 
 
In this work, a {K-A model for a mesh of curved triangles is presented}.
Its two main {features} are: (a) it is based on curvilinear triangles to take into account the curvature 
problem, (b) integrals are computed through an adaptive quadrature rule as an alternative to settle the high 
frequency ranges. 

This paper is organized as follows. Section \ref{model_d} addresses the model; the curved triangle description and 
coordinatization as well as the resulting integrals and its numerical evaluation.
In Section \ref{validation_model} the model is validated against certain known K-A exact solutions and in
Section \ref{comparison_planar} a comparison with a modern representative planar triangle model \cite{Abawi-Kirchhoff} 
(which uses exact triangle integration) is provided. In Section \ref{examples} an application to underwater acoustics is 
exhibited in order to show the relevance of the model.
The conclusions of the work are summarized in Section \ref{conclusions}.

\section{Model description}
\label{model_d}

The model calculates the far-field backscattering amplitude function $f_{\infty}$ considering the surface of the 
scattered body given by a curved-triangle mesh $\mathcal{M} = \bigcup_{l=1}^N \triangle_l $, through 
\be
	f_{\infty} = \sum_{l=1}^N \; \frac{ik}{2\pi}
	\int_{\triangle_l} \kappa_l (\vb{x}) 
	\:\mathrm{e}^{2ik\hat{k}\cdot\vb{x}} \: \hat{k}\cdot\hat{n}(\vb{x})\: dS(\vb{x}) .
	\label{fbs_total}
\ee

The condition imposed by the K-A, that the integral must be exclusively calculated over the insonified surface, is 
accomplished introducing the function $\kappa$ whose definition is  
\[
	\kappa_l (\vb{x}) = \begin{cases} 1 \quad \text{if} \:\vb{x}\; \text{is insonified} \\
	0 \quad \text{if} \:\vb{x}\; \text{is in {\it shadow}} \end{cases}
\]

With the help of this function the summatory in Eq. \eqref{fbs_total} only encompasses the contributions of the 
insonified areas in each $l$-th triangle. 

\subsection{Curved triangle description and integration}

The integration over the individual triangles of the mesh is then carried out using a description of 
second order triangles based on the shape functions $L_i$, defined by Carley \cite{Carley-quad} as
\[
\begin{aligned}
	L_1 &= 2( 1 - \xi - \eta )( \sfrac{1}{2} - \xi - \eta) \\
	L_2 &= 2 \xi( \xi - \sfrac{1}{2})  \\
	L_3 &= 2 \eta( \eta - \sfrac{1}{2}) \\
	L_4 &= 4 \xi( 1 - \xi - \eta ) \\
	L_5 &= 4 \xi \eta  \\
	L_6 &= 4 \eta ( 1 - \xi - \eta ).
\end{aligned}
\]

\begin{figure}[thb]
	\begin{center}
	\includegraphics[width=0.375\textwidth]{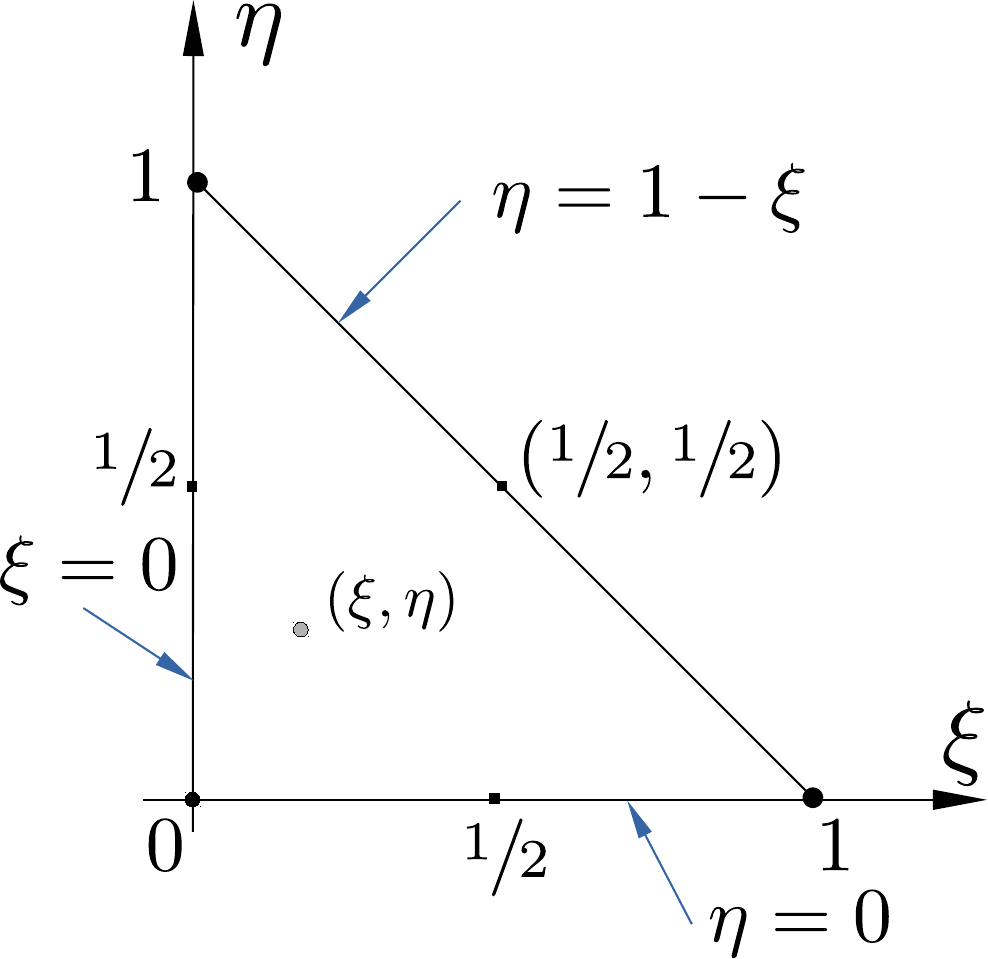} \hspace*{5mm}
	\includegraphics[width=0.55\textwidth]{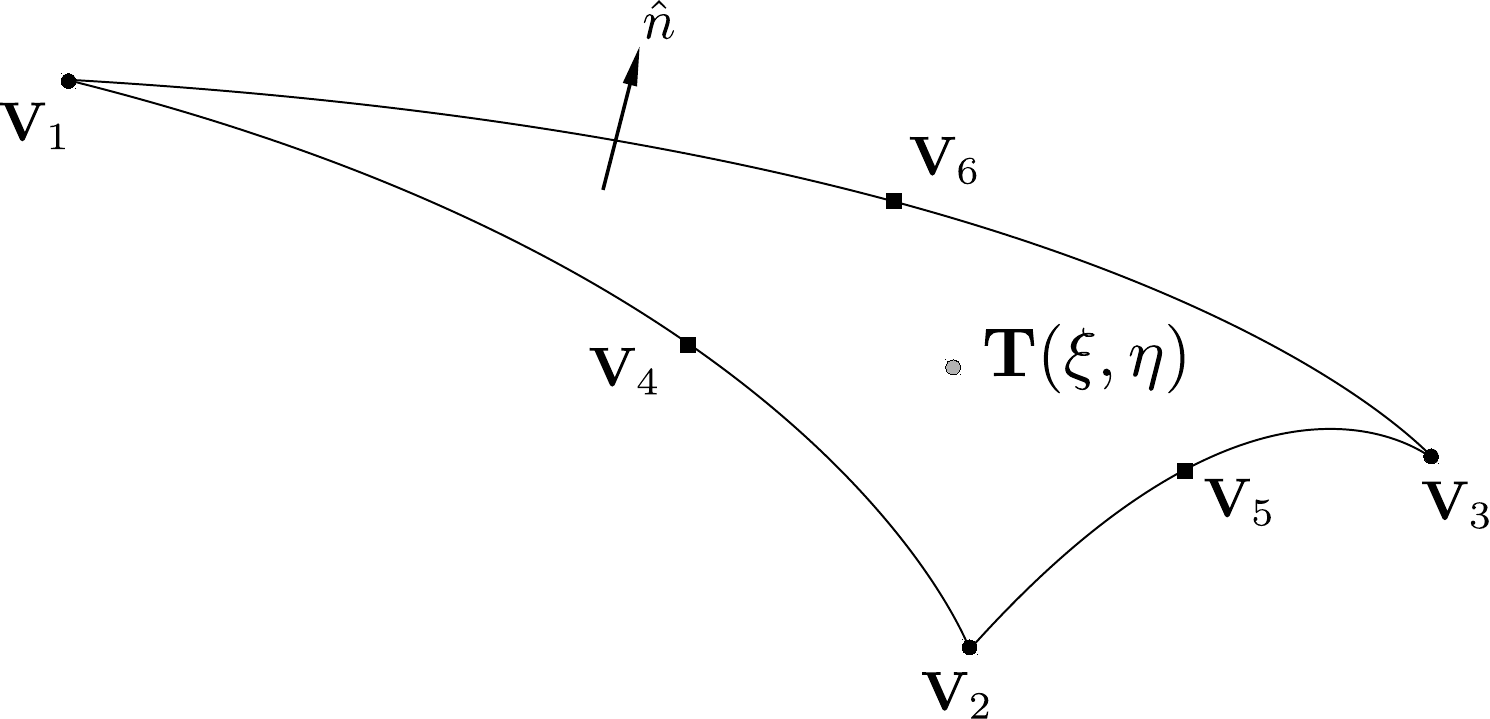}
	\end{center}
	\caption{Reference triangle (left) and curved triangle (right).}
	\label{fig_triangle_ref}
\end{figure}

These functions describe the reference triangle in terms of parameters $\xi,\eta$ that verify $0 \leq \xi \leq 1,\; 0 
\leq \eta \leq 1 - \xi$, as shown in the left side of Figure \ref{fig_triangle_ref}.
Thus, a point $\vb{T}$ belonging to the curved triangle is given by
\be
	\vb{T}(\xi,\eta) = \sum_{i=1}^6 \; L_i(\xi,\eta) \vb{V}_i
	\label{param_triangles}
\ee
where $\vb{V}_i$ are the curved triangle vertices, illustrated in Figure \ref{fig_triangle_ref} (right).

The integration over a single triangle appearing in Eq. \eqref{fbs_total} in terms of the parameterization 
\eqref{param_triangles} reads
\[
	\int_\triangle \kappa_l(\vb{x})\: \euler{2ik\hat{k}\cdot\vb{x}} \: 
	\hat{k}\cdot\hat{n}(\vb{x})\: dS(\vb{x}) =
	\int_0^1 \int_0^{1-\xi} \kappa_l(\vb{T})\: \euler{2ik(\hat{k}\cdot\vb{T})} \;
	\hat{k}\cdot ( \vb{T}_\xi \times \vb{T}_\eta ) \; d\eta d\xi
\]
where 
\[
	\hat{k}\cdot\vb{T} =  C_1\xi^2 + C_2\xi + C_3\eta\xi + C_4\eta^2 + C_5\eta + C_6
\]	
\[
	\hat{k}\cdot ( \vb{T}_\xi \times \vb{T}_\eta ) = 
	D_1\xi^2 + D_2\xi + D_3\xi\eta + D_4\eta^2 + D_5\eta + D_6,
\]
being $C_i$ and $D_i$ constants defined in terms of the $\vb{V}_i$ ($i=1,2,...,6$). 
These expressions are tabulated {explicitly} in the Appendix.

Now, the contribution $f_{\infty}^{\triangle_l}$ from the $l$-th triangle to the total $f_{\infty} = 
\sum_l^N f_{\infty}^{\triangle_l} $ is
\[
	f_{\infty}^{\triangle_l} =  \frac{ik}{2\pi}\:\euler{2ikC_6} \: F_l(\{C_m\},\{D_n\})
\] 
where 
\begin{multline}
	F_l(\{C_m\},\{D_n\}) = \int_0^1 \: \euler{2ikP(\xi)}
	\left[ (D_1\xi^2 + D_2\xi + D_6) \int_0^{1-\xi} \kappa_l(\vb{T}) \: \euler{2ik Q(\xi,\eta)} d\eta \; + 
	\right. \\ 
	\left. (D_3\xi + D_5) \int_0^{1-\xi} \kappa_l(\vb{T}) \: \euler{2ik Q(\xi,\eta)} \eta d\eta  +
	 D_4 \int_0^{1-\xi} \kappa_l(\vb{T}) \: \euler{2ik Q(\xi,\eta)} \eta^2 d\eta \right] d\xi
\end{multline}
and
\[
	P(\xi) = C_1\xi^2 + C_2\xi, \qquad \qquad \qquad \qquad Q(\xi,\eta) = C_3\eta\xi + C_4\eta^2 + C_5\eta.
\]

The function $F(\{C_m\},\{D_n\})$ ($m=1\dots 5,n=1\dots 6$) accounts for the three iterated integrals over $\eta$ and 
$\xi$.
These integrals are calculated through a two nested quadrature rule implemented by a specialized computational routine 
QuadGK \cite{quadgk} available under the Julia language programming \cite{julia}.

In summary, the model consists of the following three key elements: (1) a curved triangle mesh $\mathcal{M}$, (2) a 
function $\kappa$ that determinates the insonified surface and (3) an integration algorithm over a curved triangle based 
on iterated quadrature rules.

\section{Validation}
\label{validation_model}

We validate the model against the K-A for three simple shapes whose backscattering amplitude function $f_\infty$ 
has an analytical closed-form solution; the sphere, the cylinder with flat end caps and the prolate spheroid.
In each case we present plots of $ |f_{\infty}| $ in terms of an adimensional parameter $ k\ell $ (being $\ell$ a 
characteristic length), which is the usual magnitude of interest in many applications.

\subsection{Sphere}

For a sphere of radius $r$, the K-A integral has a simple closed-form \cite{Abawi-Kirchhoff}, 
\be
	f_{\infty} = \frac{i}{4k} \: \euler{-2ikr} \left( \euler{2ikr} - 2ikr - 1 \right).
	\label{ka_sphere}
\ee

\begin{figure}[hbt]
	\begin{center}
	\includegraphics[width=0.95\textwidth]{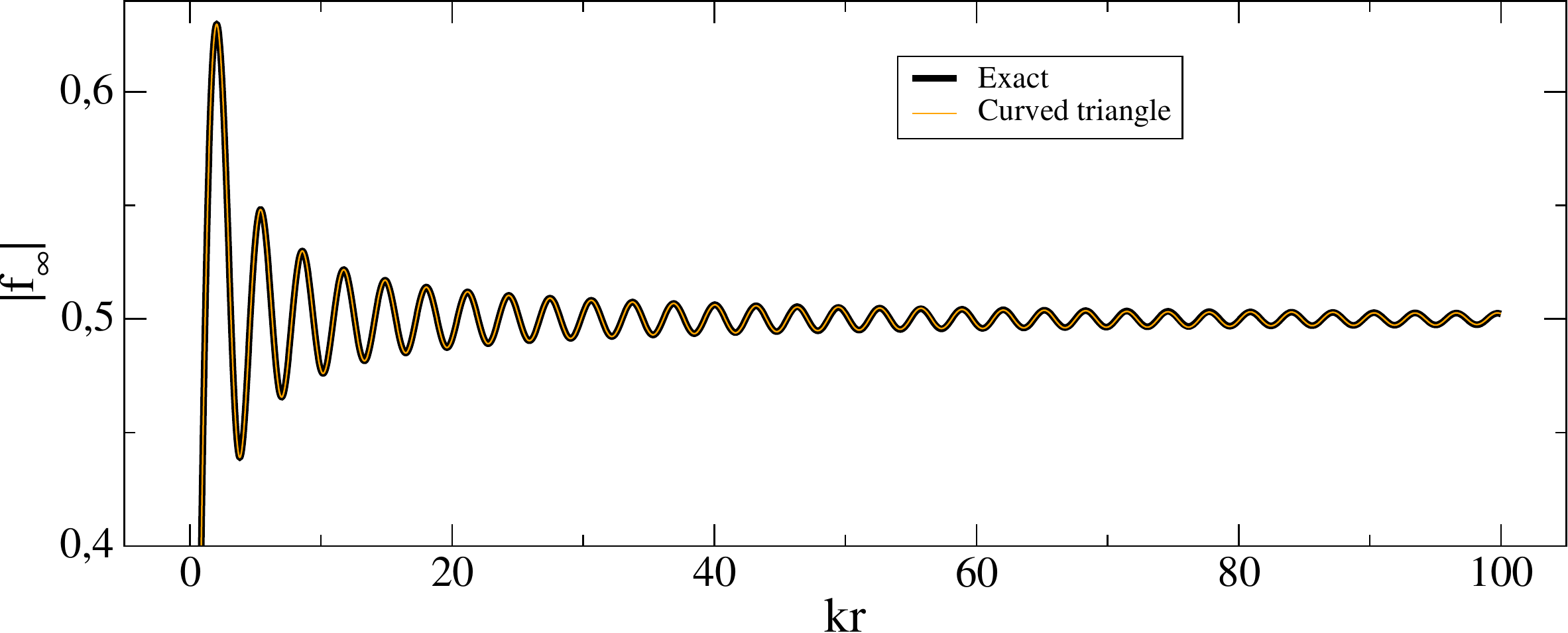}
	\end{center}
	\caption{Results for the $|f_{\infty}|$ of the sphere. The mesh used for the curved triangle model has 1278 
	triangles.}
	\label{fig_sphere_bs}
\end{figure}

For the comparison with the exact solution, a mesh with 1278 curved triangles has been used. The resulting 
$|f_{\infty}|$ is exhibited in Figure \ref{fig_sphere_bs} for the interval $0.1 \leq kr \leq 100$ where $r=1$ is the 
sphere {radius}.

\subsection{Cylinder with flat endcaps}

\begin{figure}[H]
	\begin{center}
	\includegraphics[width=0.27\textwidth]{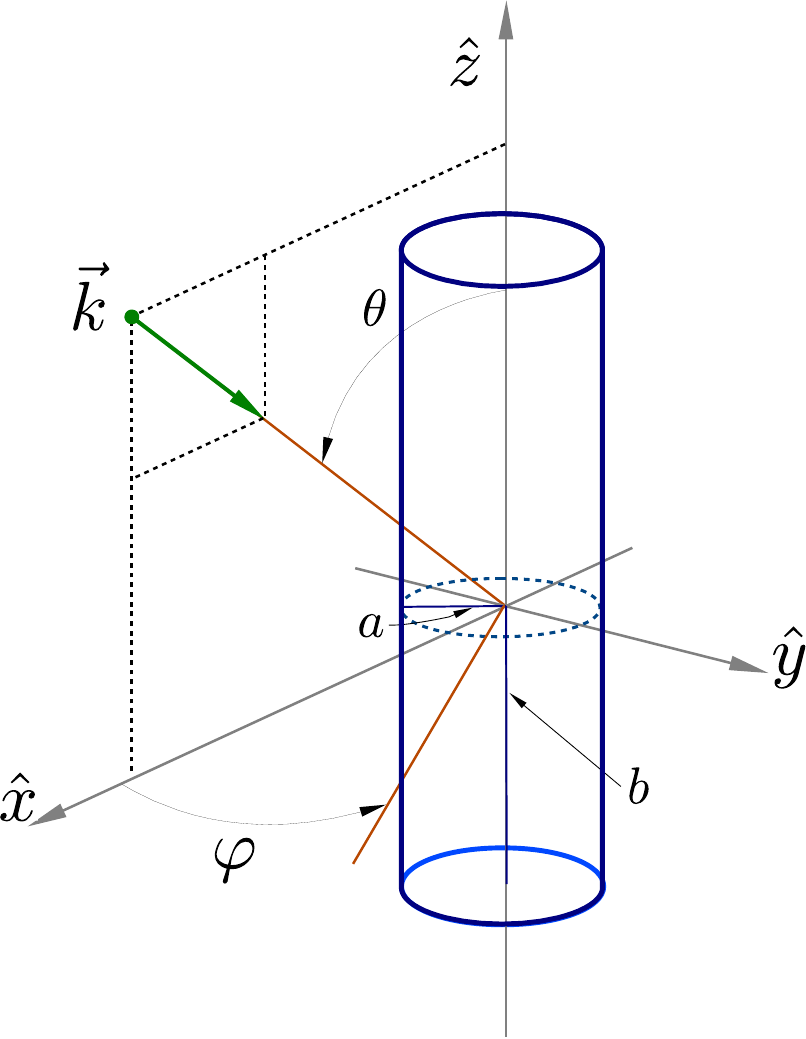} \hspace*{15mm}
	\includegraphics[width=0.27\textwidth]{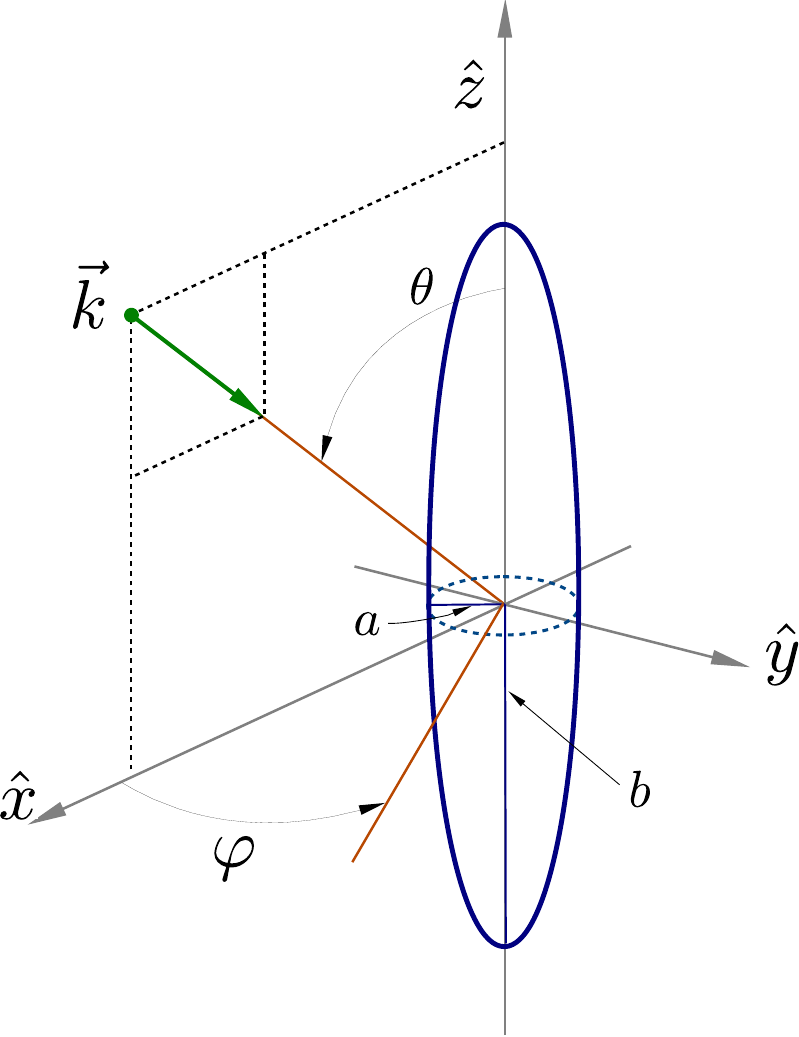} \hspace*{15mm}
	\end{center}
	\caption{Schemes for the incidence over the cylinder with flat endcaps (left) and the spheroid (right).
	The dimensions used in all {the} tests were $a=0.5$ and $b=1$.}
	\label{fig_schemes}
\end{figure}

The geometry and the coordinates considered for the insonification of the {following} validation examples, 
namely, cylinder and spheroid, are shown in Figure \ref{fig_schemes}.

The expression of the backscattering from a finite cylinder of radio $a$ and length $2b$ (without endcaps) for an 
incidence direction characterized by the $\theta$ angle is \cite{Gaunard},
\be
	f_{\infty}^{cyl} = -\frac{a}{2} \tan(\theta) \sin(2kb\cos(\theta))\left[ \frac{2}{\pi} - 
	\mathbf{H}_1(2ka\sin(\theta)) - iJ_1(2ka\sin(\theta))\right],
	\label{cyl_lateral}
\ee
where $\mathbf{H}_1$ and $J_1$ are the first order Struve and cylindrical Bessel functions, respectively.
The contribution of the top flat surface ({circle of radius $a$}) is
\be
	f_{\infty}^{top} = -ika^2 \cos(\theta)\: \euler{-2ikb\cos(\theta)} \:
	\frac{J_1(2ka\sin(\theta))}{2ka\sin(\theta)}.
	\label{flat_encap_cyl}
\ee

In order to obtain the exact K-A solution for the cylinder is enough to consider the sum of the expressions of Eqs. 
\eqref{cyl_lateral} and \eqref{flat_encap_cyl}.
These account for the backscattering in the $ 0 \leq \theta \leq \pi/2 $ incidence while the range 
$ \pi/2 \leq \theta \leq \pi $ is easily built up from the former using symmetry considerations.

\begin{figure}[!h!bt]
	\begin{center}
	\includegraphics[width=0.90\textwidth]{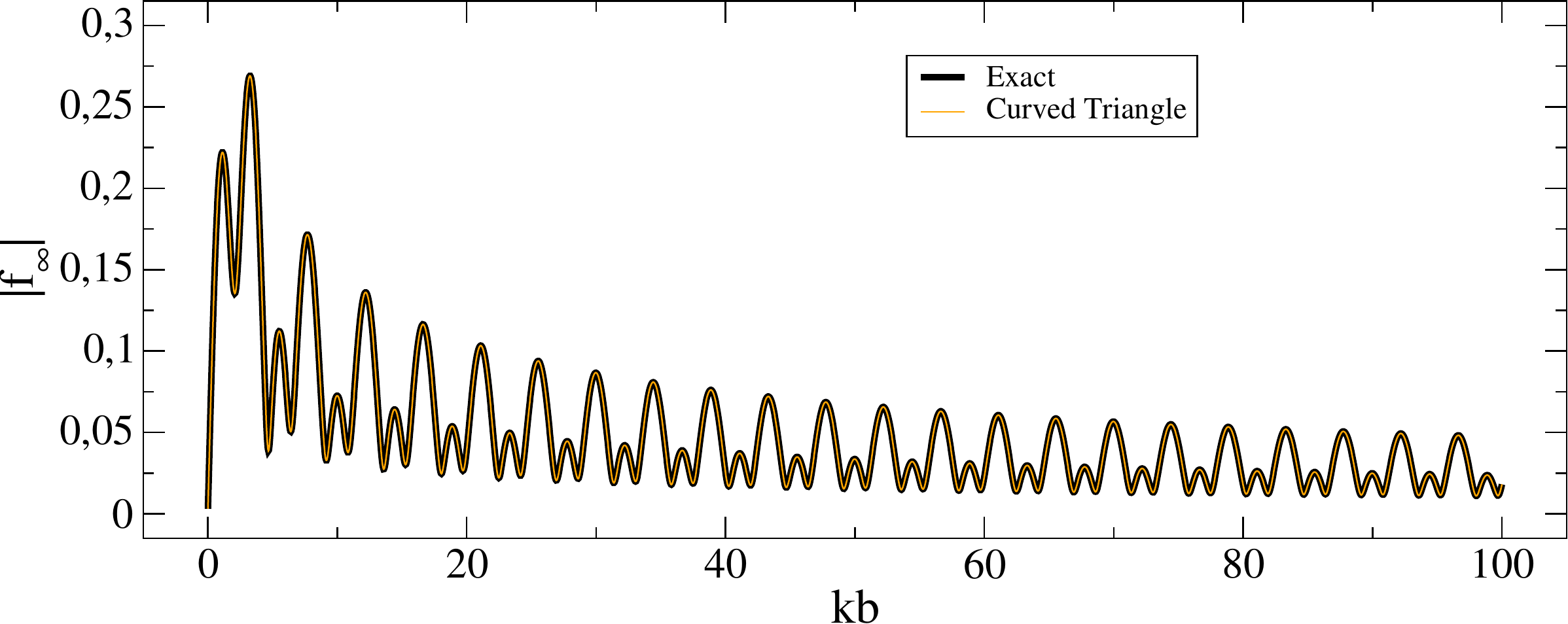}
	\end{center}
	\caption{Results for the $|f_{\infty}|$ of the flat endcaps cylinder (radius $a=0.5$, semilength $b=1$) at 
	incidence angle $\theta = \pi/4$. The mesh used for the curved triangle model has 1238 triangles.}
	\label{fig_cylinder_bs}
\end{figure}

The agreement between the exact solution and the curved triangle model for the test mesh ($a=0.5$ and $b=1$) is 
illustrated in Figure \ref{fig_cylinder_bs}. The mesh consists of 1238 curved triangles. 
The fixed incidence used was $\theta = \pi/4$.

\subsection{Prolate spheroid}

Considering a prolate spheroid of semiaxes $a$ and $b$ ($b > a$), the backscattering amplitude function $f_\infty$ 
can be obtained in a closed-form for the cases $\theta=0$, i.e. parallel to the foci line, and $\theta=\pi/2$, 
perpendicular to the foci line, (see Figure \ref{fig_schemes}, right).
In these cases we have 
\be
	f_\infty(\theta\!=\!0) = -\frac{i}{4k} \left(\frac{a}{b}\right)^2 \euler{-2ikb}
	\left( \euler{i2kb} - 2ikb - 1 \right)
	\label{ka_spheroid_theta_0}
\ee
and
\be
	f_\infty(\theta\!=\!\pi/2) = -\frac{i}{4k} \left(\frac{b}{a}\right) \euler{-2ika}
	\left( \euler{i2ka} - 2ika - 1 \right),
	\label{ka_spheroid_theta_piover2}
\ee
both expressions having the same form that Eq. \eqref{ka_sphere}.
In particular, at an incidence angle $\theta=0$ the Eq. \eqref{ka_spheroid_theta_0} verifies 
\[
	f_\infty^{spheroid} = (a/b)^2 f_\infty^{sphere}(r=b),
\]
i.e. the backscattering corresponds to a multiple of the backscattering from a sphere.

\begin{figure}[!h!bt]
	\begin{center}
	\includegraphics[width=0.9\textwidth]{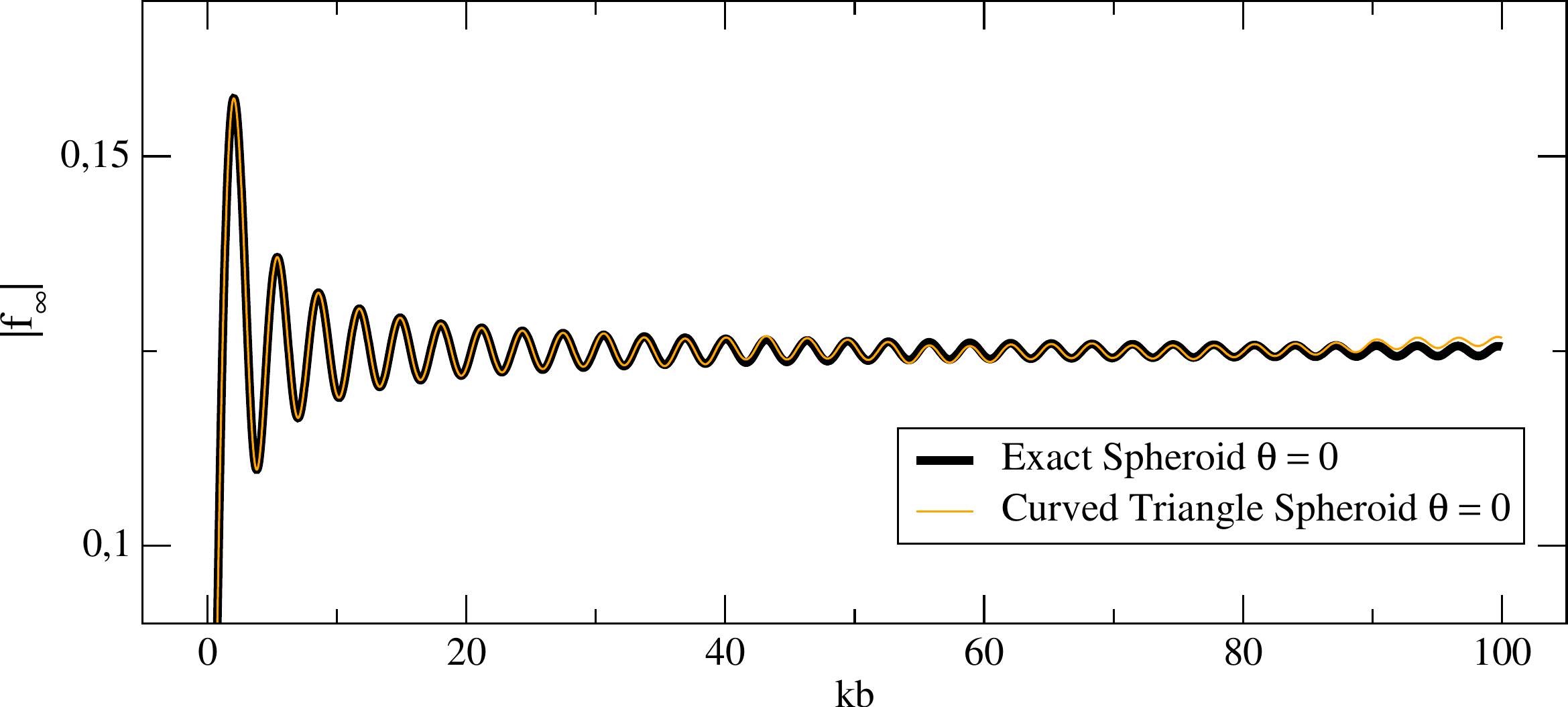}
	\end{center}
	\caption{Results for the $|f_\infty|$ of the prolate spheroid at the incidence $\theta=0$.
	The mesh used for the curved triangle model have 1386 curved triangles.}
	\label{fig_spheroid_bs_0}
\end{figure}

\begin{figure}[!h!tb]
	\begin{center}
	\includegraphics[width=0.9\textwidth]{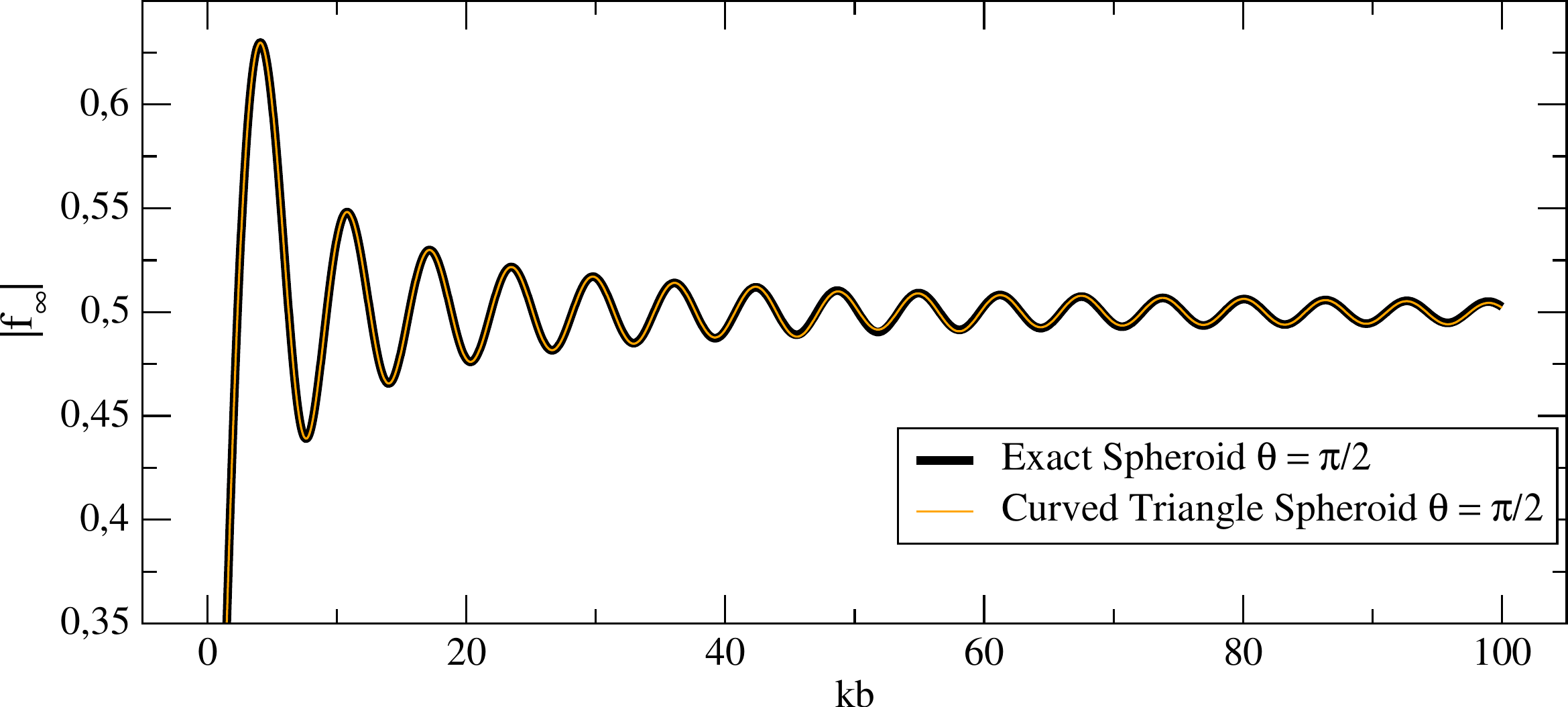}
	\end{center}
	\caption{Results for the $|f_\infty|$ of the prolate spheroid at the incidence $\theta = \pi/2$.
	The mesh used for the curved triangle model have 1386 curved triangles.}
	\label{fig_spheroid_bs_pi2}
\end{figure}

The corresponding $| f_\infty |$ at the incidences $ \theta = 0 $ and $ \theta = \pi/2 $ are shown in Figure 
\ref{fig_spheroid_bs_0} and Figure \ref{fig_spheroid_bs_pi2}, respectively. In both cases a spheroid mesh of 1386 
triangles was used and the $|f_\infty|$ is plotted against $ kb $, being $ b=1 $ the major semiaxis.

Some errors are visually evident in the Figure \ref{fig_spheroid_bs_0} which corresponds to the $ \theta = 0 $ incidence.
Here the incident wave faces the spheroid in the end-on aspect, so that curvature issues are more crucial than in beam 
aspect ($ \theta = \pi/2 $). Evidently for $ kb > 90 $ a more refined mesh is neccessary.

\section{Comparison with planar triangles}
\label{comparison_planar}


When using the flat triangles the integration results in a expression simpler than on any other type of curved 
surface but as drawback artificial edges are introduced and, as stated in Section \ref{intro}, a big number of 
triangles are necessary for describing a general geometry with negligible error in curvature unless the obstacle is a 
polyhedron.

The planar triangle model consequently requires for the correct description of curved surfaces a much bigger mesh than 
in the curved case.
In a recent work \cite{Abawi-Kirchhoff} the author recommends that the all triangle's edge size $e$ must verify the 
relation $10e < \lambda$ (being $\lambda=2\pi/k$ the wavelength) for an accurate backscattering evaluation (ten elements
per wavelength).
Moreover, because of the flatness of the planar triangles, the determination of the insonified area is also approximate.
Each triangle has a unique normal $\hat{n}$ so that a given triangle can only be fully insonified or fully in shadow. 

To evaluate the performance of both approaches we take a unit ($r=1$ radius) sphere mesh with
$ N = 42472, 171472, 982874 $ planar triangles and a curved one of $ N = 2016 $.
For the range $0.1 \leq kr \leq 100$ we have evaluated the $f_\infty$ and the absolute value of the relative error,
defined as 
\be
	\frac{|f_\infty^{model} - f_\infty^{exact}|}{|f_\infty^{exact}|}.
	\label{rel_error}
\ee

\begin{figure}[htb]
	\begin{center}
	\includegraphics[width=0.9\textwidth]{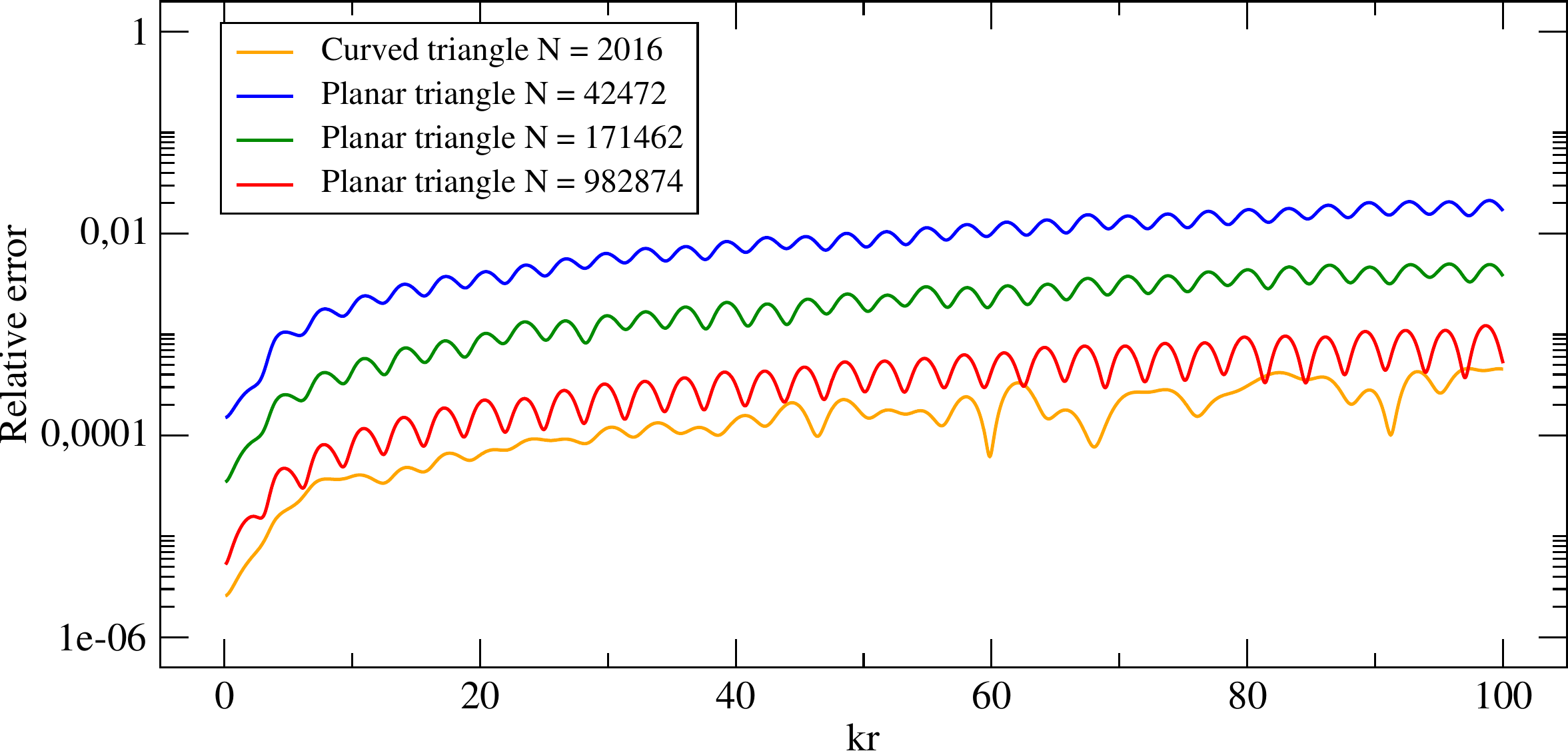}\vspace*{-5mm}
	\end{center}
	\caption{Relative error in the $f_\infty$ according to \eqref{rel_error} in the backscattering 
	calculation for a unit radius sphere with a planar triangle mesh ($ N = 42472 $, $ N = 171472 $ and $ N = 
	982874 $) against a curved triangle mesh of $ N = 2016 $.}
	\label{fig_sphere_error}
\end{figure}

To avoid biased results caused by anisotropies of the mesh, the amplitude function $f_\infty$ presented here is
the average of three incidences along the coordinate axes.
The relative error curves for this comparison are shown in Figure \ref{fig_sphere_error}. 
It is evident that the curved triangle model obtains better fit (minor relative error) in the entire frequency range 
considered, even against a planar triangle mesh with a number of triangles $\sim 500$ times larger.

To test both schemes in high frequency we again calculate the relative error (as a average of three incidences) in the 
case of two plane triangle meshes of $ N = 5034, 6081410 $ and a curved one of $ N = 5034 $ in the range $0.1 < kr < 1000$ 
($kr=1000$ stands for a sphere with a perimeter of 159$\lambda$).

\begin{figure}[!h!bt]
	\begin{center}
	\includegraphics[width=0.9\textwidth]{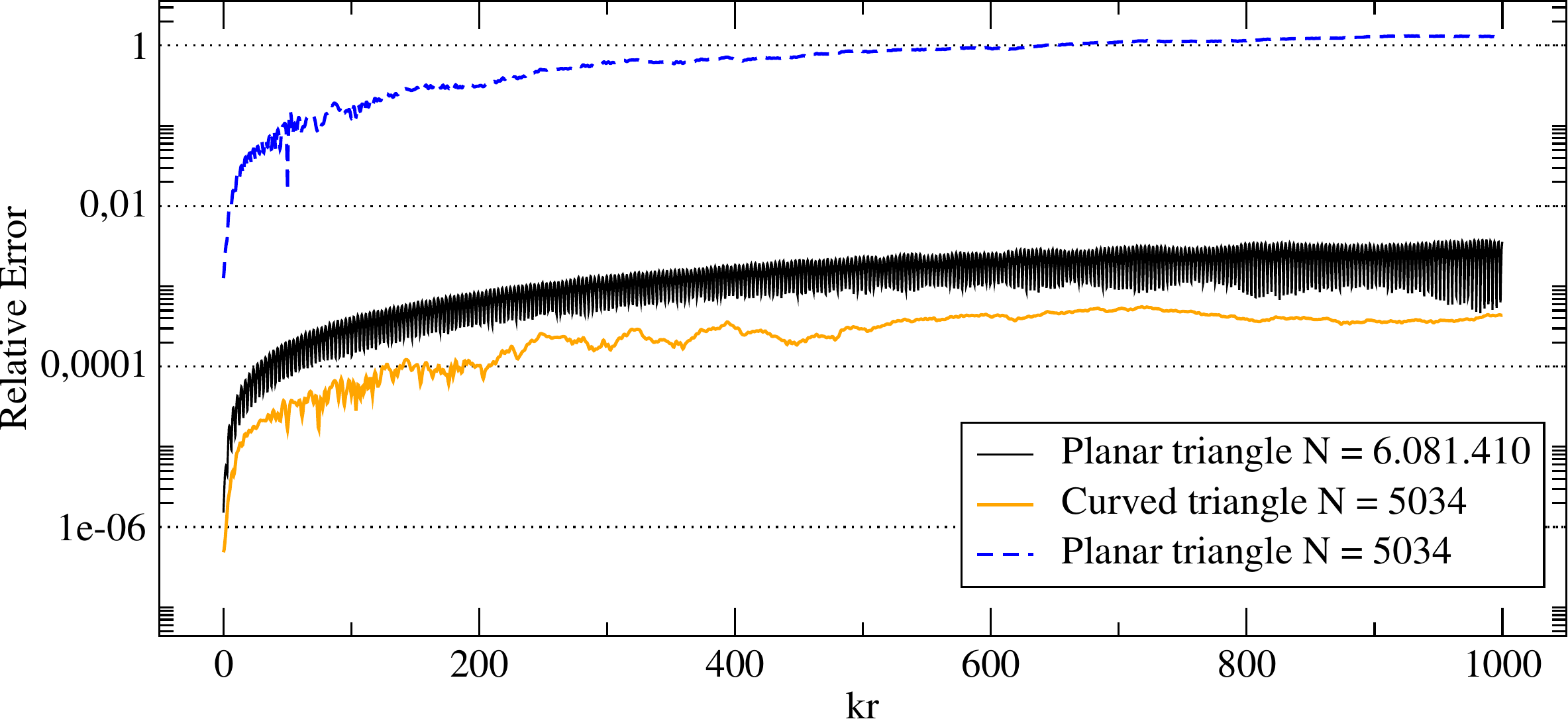}\vspace*{-5mm}
	\end{center}
	\caption{Relative error in the $f_\infty$ according to \eqref{rel_error} in the backscattering calculation for 
	a unitary sphere under two planar triangle meshes of $N=5034, 6081410$ against a curved triangle mesh of $N=5034$.}
	\label{fig_sphere_vhf_error}
\end{figure}

Keeping constant the number of triangles ($N=5034$) the curved mesh achieves a maximum relative error, in the entire range, of 
$1/2000$ while in the planar mesh it rises to $ 1.3 $. For the plane mesh $N = 6081410$, three orders of magnitude bigger than
the curved one, the relative error was yet greater as is illustrated by the Figure \ref{fig_sphere_vhf_error}.


\section{Application example}
\label{examples}

The K-A constitutes a common basis on which several algorithms for the numerical calculation of the scattering by 
submarines are developed \cite{ts_subs,improved_betssi}.
The backscattering produced by submarines, relevant in the field of underwater acoustics, is usually expressed in terms 
of the {\it Target Strength} TS \cite{MedwinClay}, defined as
\[
	\text{TS} = 20 \log_{10} ( |f_\infty| ).
\]
In this section, aimed to applications, we use the curved triangle model to evaluate the TS for the backscattering of a 
submarine.

\begin{figure}[H]
	\begin{center}
	\includegraphics[width=0.6\textwidth]{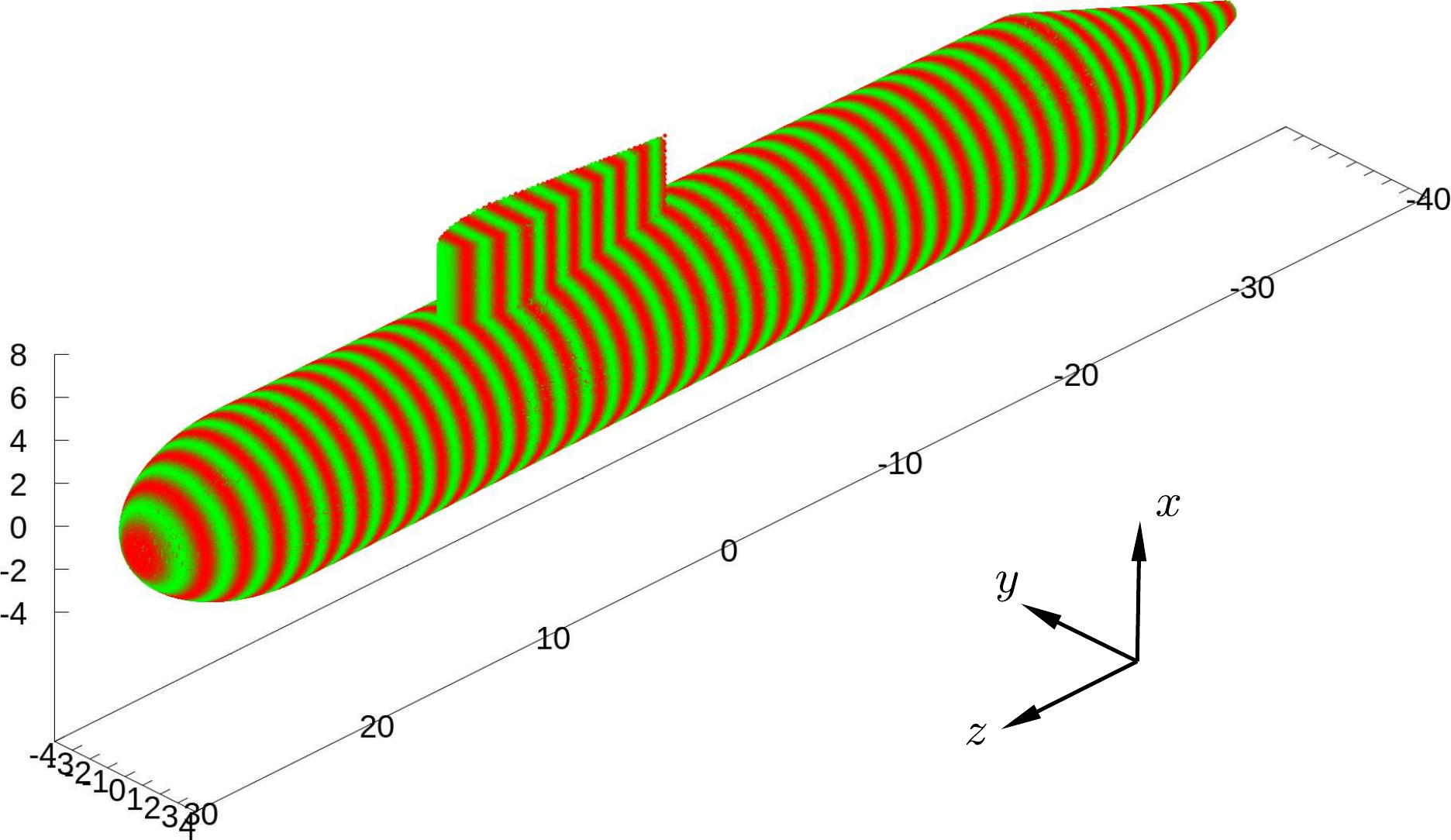}\vspace*{-5mm}
	\end{center}
	\caption{Generic submarine used for testing.
	The real part of a plane wave of $f=1000$ Hz and incidence direction given by $-\hat{z}$ 
	is showed on the submarine surface.}
	\label{fig_submarin0_simple}
\end{figure}

To test our curved triangle model we construct a generic simplified submarine (based on the simple BeTSSi model 
\cite{improved_betssi}), which is shown in Figure \ref{fig_submarin0_simple}.
From this submarine two meshes are built; a planar one with $N=77220$ triangles and a curved one with $N=18116$.
The planar mesh is used to evaluate a reference solution for the backscattering problem which is based on a BEM method.

The length $ \ell $ of the triangle's side (edges) for the plane mesh has a mean value of $0.207$ m and verifies the
condition $ \lambda \geq 6 \ell $ for 93 \% of the ocurrences (the length distribution is approximately gaussian).
This relation between edge length $\ell$ and wavelength $ \lambda $ assures that the object is adequatelly represented 
by the mesh in terms of the backscattering calculation for the BEM reference solution.

The BEM method, under a conventional implementation, provides relatively accurate solutions only for the low or 
intermediate frequency regimes. Nevertheless it's expected that the K-A coincides with this reference solution (for 
some incidences) even in the intermediate frequency regime because the conditions of the former are satisfied under 
several incidences in a object like the actual test submarine.

The generic submarine can be characterized by two lengths; those corresponding to the length of the cylinder that 
forms the {external} hull and its radius. These lengths are $ a = 31 $ m and $ b = 3.75 $ m.
For a incident wave of frequency $f$ the corresponding wave number is $ k = 2\pi f/ c $ where $c$ is the sound speed in 
the water. For the case $f = 1000$ Hz and taking $ c = 1500 $ m/seg (an accepted average value) we have $ k = 4.188 $ 
and the scattering can be adimensionally characterized by $ ka \approx 130 $ and $ kb \approx 16 $. The Figure 
\ref{fig_submarin0_simple} shows the real part of an incident wave $\exp({i k \hat{k}\cdot\vb{x}})$ 
(with $\hat{k}=(0,0,-1)$) evaluated on the submarine surface.

\begin{figure}[tbh]
	\begin{center}
	\includegraphics[width=0.85\textwidth]{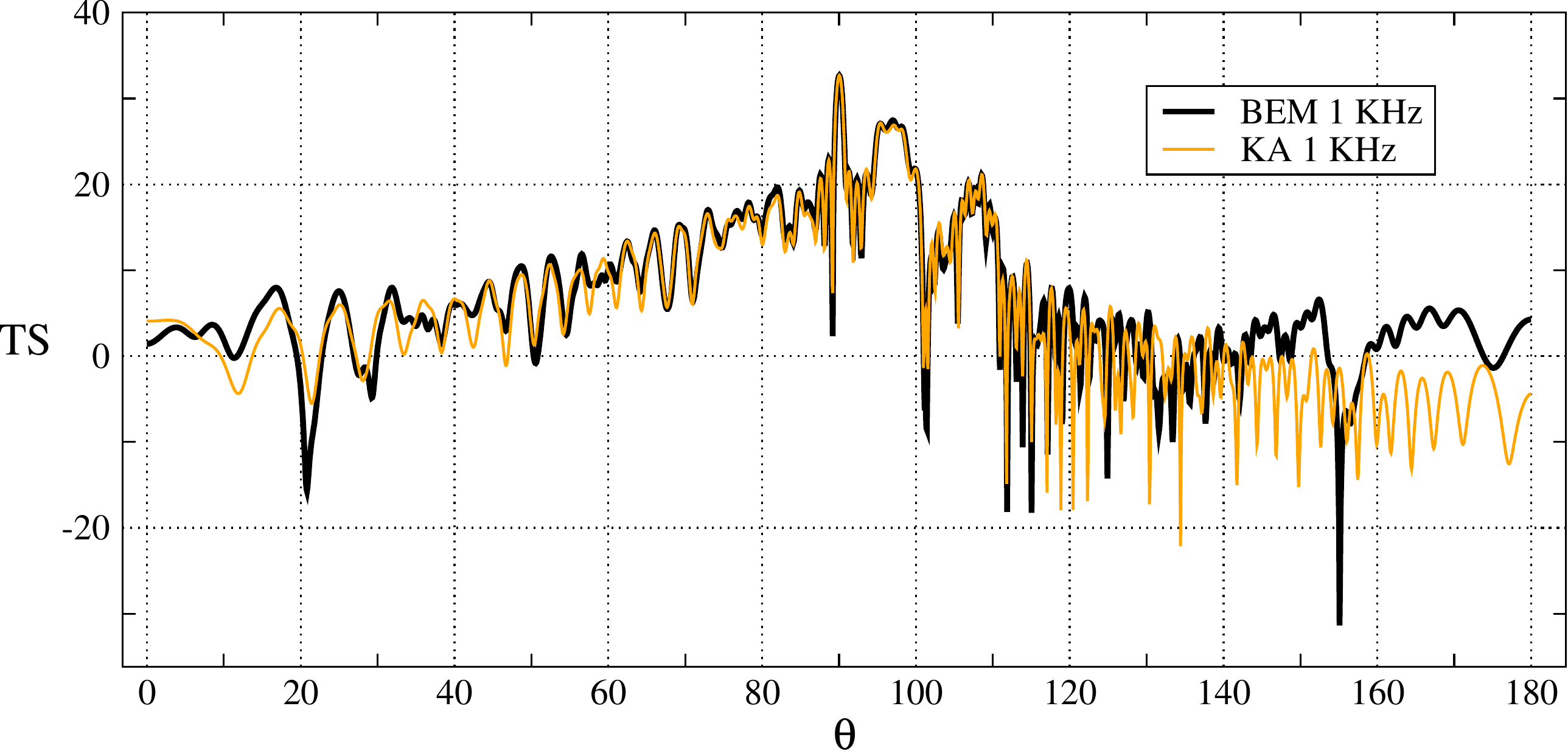}\vspace*{-5mm}
	\end{center}
	\caption{TS for the submarine model for an incidence $ \hat{k} = -(0,\sin \theta,\cos \theta)$ and frequency 
	$f = 1000$ Hz evaluated by the K-A with a curved triangle mesh and by a BEM solution (used as a reference).}
	\label{fig_ts_sub}
\end{figure}

For the TS calculation we restrict ourselves to incidence directions with altitude 0$^\circ$ (i.e. incidence belonging 
to the $ yz $ plane), parameterized in terms of a $\theta$ angle by $\hat{k} = -(0, \sin\theta, \cos \theta) $ .
The results of the backscattering TS for the submarine model at $ 0^\circ $ altitude is shown in Figure \ref{fig_ts_sub}
for the K-A with the curved mesh and the BEM reference solution with the planar one for the frequency $f=1$ KHz. 
It follows from the figure that the K-A provides a good agreement, specially up to $115^\circ$.

\begin{figure}[bth]
	\begin{center}
	\includegraphics[width=0.85\textwidth]{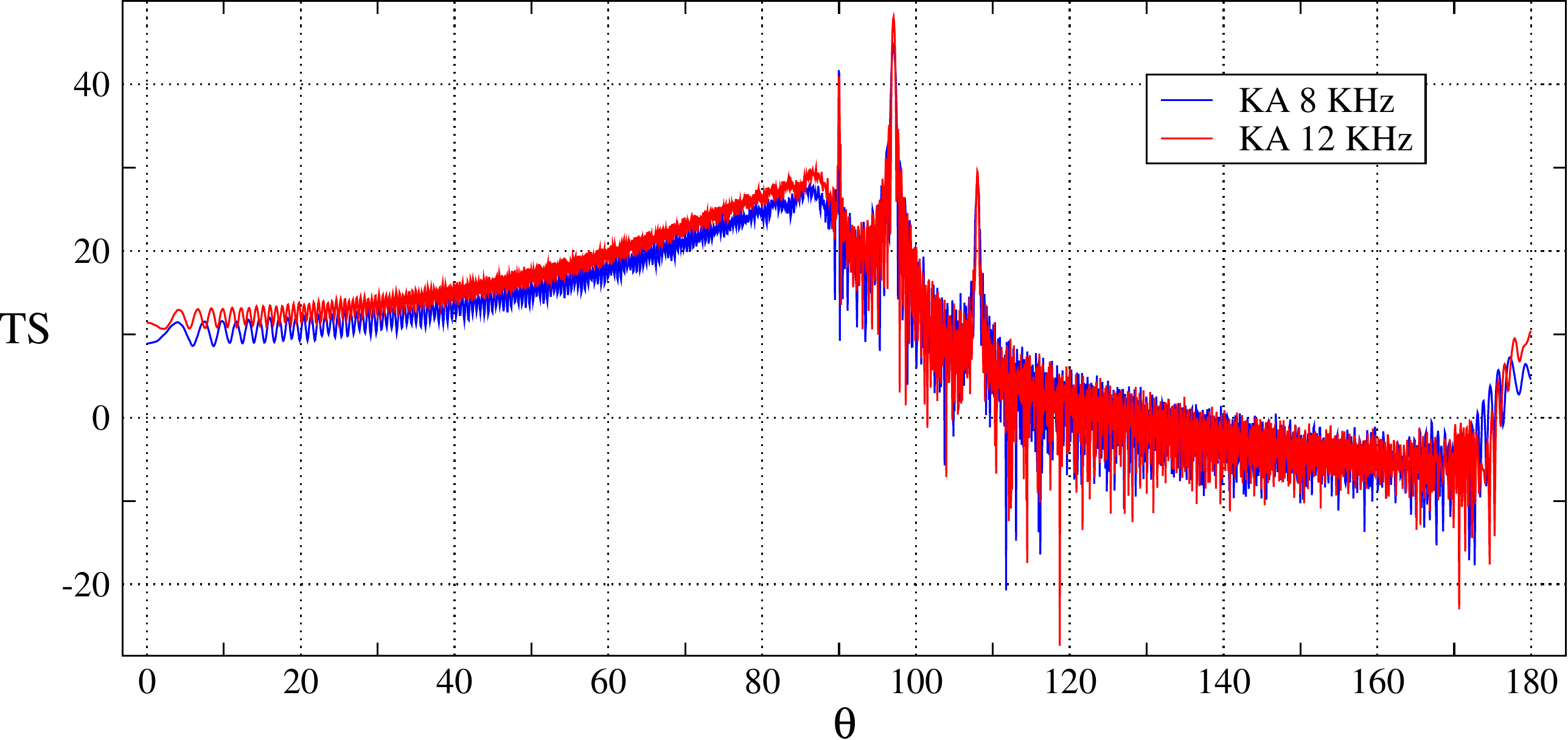}\vspace*{-5mm}
	\end{center}
	\caption{TS for the submarine model for an incidence $ \hat{k} = -(0,\sin \theta,\cos \theta)$ and frequencies 
	$f = 8$ KHz and $f = 12$ KHz evaluated by the K-A with the curved triangle mesh.}
	\label{fig_ts_sub_vhf}
\end{figure}

It is expected that for $\theta$ tending to $180^\circ$ (stern direction) the K-A validity ({as solution of the
backscattering problem}) weakens because in these incidences the submarine presents a less smooth surface due to 
the presence of the edges that form the union of the aft (which in this simplified model is a cone) with the main body 
and those that constitutes the back of the veil.
Any effects due to edge's diffraction are, of course, not taken into account in the K-A.

The TS for $ 0^\circ $ altitude and the frequencies $f=8$ KHz and $f=12$ KHz which means for $ ka \approx 1039, 
kb \approx 126 $ and $ ka \approx 1558, kb \approx 188 $, respectively, is shown in the Figure \ref{fig_ts_sub_vhf}.

\section{Conclusions}
\label{conclusions}

The model presented in this work, based on numerical iterated integration over curved triangles and with exact 
accounting of the insonified surface of the scatterer, has been able to provide excellent agreement against
known exact solutions of the K-A (the validation examples of section \ref{validation_model}) up to high $ k \ell $ 
values and requiring only small size meshes.
The application's range previously reported in the literature could be also extended.

The comparison with a model based on flat triangle integration, performed in section \ref{comparison_planar}, showed
that a small mesh of curved triangles can provide better fit than a much larger mesh of planar ones.


While the curved triangle model is somewhat complex because requires numerical integration techniques, it has the 
advantage of adequately modeling curved surfaces with a small mesh size, and also determine in a more precisely way 
the insonified area due to triangles that can be partially insonified, in contrast to the case of planar triangular 
facets.
In the latter, although the integration is almost trivial the only way to overcome the curvature and partial 
insonification issues is through the increase in mesh size and that will lead, sooner or later, to a bottleneck in 
computational storage if the frequency is high enough.

On the other hand, the algorithm for the calculation of backscattering from curved triangles is easily parallelizable 
so that it can take advantage of the existence of multiple cores in the computer where it is executed.
This fact can contribute significantly to decrease the computing time.

The model can provide an approximate technique for the determination of complex object backscattering in the high 
frequency regime as it has been shown in the section \ref{examples}, where the backscattering of a generic submarine 
was calculated under this approach and good agreement (compared to a reference solution) was obtained for incidence 
directions over which the K-A applicability was guaranteed.

\clearpage
\appendix

\section{Constants $C_i$ $D_i$}

Expressions of the constants involved in the integration over the curved triangles in terms of the vertices $\vb{V}_i$.
\[
\begin{aligned}
	C_1 &= \hat{k}\cdot( \vb{V}_1 + \vb{V}_2 - 2\vb{V}_4 ) \\
	C_2 &= \hat{k}\cdot( -3\vb{V}_1 - \vb{V}_2 + 2\vb{V}_4 ) \\
	C_3 &= \hat{k}\cdot( \vb{V}_1 - \vb{V}_4 + \vb{V}_5 - \vb{V}_6 ) \\
	C_4 &= \hat{k}\cdot( \vb{V}_1 + \vb{V}_3 - 2\vb{V}_6 ) \\
	C_5 &= \hat{k}\cdot( -3\vb{V}_1 - \vb{V}_3 + 4\vb{V}_6 ) \\
	C_6 &= \hat{k}\cdot \vb{V}_1   
\end{aligned}
\]

\[
\begin{aligned}
	D_1 = 16\;\left( \hat{k}\times\vb{V}_4\cdot\left[-\vb{V}_1 + \vb{V}_2 - 2\vb{V}_5 + 2\vb{V}_6\right] +
	\hat{k}\times\vb{V}_2\cdot\left[ \vb{V}_1 + \vb{V}_5 - \vb{V}_6 \right] +
	\hat{k}\times\vb{V}_1\cdot\left[ \vb{V}_5 - \vb{V}_6 \right] \right)
\end{aligned}
\]
\begin{multline*}
	D_2 = -4\;\left( \hat{k}\times\vb{V}_1\cdot\left[-4\vb{V}_2 + \vb{V}_3 + 7\vb{V}_4 + 3\vb{V}_5 -
	7\vb{V}_6 \right] + \hat{k}\times\vb{V}_2\cdot\left[ \vb{V}_3 - \vb{V}_4 + \vb{V}_5 - 5\vb{V}_6 
	\right] \right.+ \\
	\left. \hat{k}\times\vb{V}_4\cdot\left[ -2\vb{V}_3 - 4\vb{V}_5 + 12\vb{V}_6 \right] \right)
\end{multline*}

\[
\begin{aligned}
	D_3 = 16\;\left( \hat{k}\times\vb{V}_1\cdot\left[ \vb{V}_3 - \vb{V}_2 + 2\vb{V}_4 - 2\vb{V}_6\right] +
	\hat{k}\times\vb{V}_2\cdot\left[ \vb{V}_3 - 2\vb{V}_6 \right] +
	\hat{k}\times\vb{V}_1\cdot\left[ -2\vb{V}_3 + 4\vb{V}_6 \right] \right)
\end{aligned}
\]

\[
	D_4 = -16\;\left( \hat{k}\times\vb{V}_1\cdot\left[-\vb{V}_3 - \vb{V}_4 + \vb{V}_5 + \vb{V}_6 \right] + 
	\hat{k}\times\vb{V}_2\cdot\left[ -\vb{V}_4 + \vb{V}_5 \right] +
	\hat{k}\times\vb{V}_6\cdot\left[ \vb{V}_3 + 2\vb{V}_4 - 2\vb{V}_5 \right] \right)
\]

\begin{multline*}
	D_5 = 4\;\left( \hat{k}\times\vb{V}_6\cdot\left[-4\vb{V}_5 + 12\vb{V}_4 - 2\vb{V}_2 - 7\vb{V}_1 
	\right] \right. + \\
	\hat{k}\times\vb{V}_1\cdot\left[ 3\vb{V}_5 - 7\vb{V}_4 - 4\vb{V}_3 + \vb{V}_2 \right] + \\
	\left. \hat{k}\times\vb{V}_3\cdot\left[ \vb{V}_2 - 5\vb{V}_4 + \vb{V}_5 -\vb{V}_6 \right] \right)
\end{multline*}

\[
	D_6 = \hat{k}\times\vb{V}_6\cdot\left[-16\vb{V}_4 + 4\vb{V}_2 + 12\vb{V}_1 \right] + 
	\hat{k}\times\vb{V}_1\cdot\left[ 12\vb{V}_4 - 3\vb{V}_2 \right] +
	\hat{k}\times\vb{V}_3\cdot\left[ -3\vb{V}_1 + 4\vb{V}_4 - \vb{V}_2 \right]
\]


\begin{thebibliography}{0}

\bibitem{MedwinClay} H. Medwin \& C.S. Clay, {\it Fundamentals of Acoustical Oceanography}, Academic 
Press, 1998.

\bibitem{Fawcett} J.A. Fawcett, {\it Modeling of high-frequency scattering from objects using a hybrid 
Kirchhoff/diffraction approach}. The Journal of the Acoustical Society of America, 109(4), 1312-1319, 2001.

\bibitem{Lee} Lee, K., \& Seong, W., {\it Time-domain Kirchhoff model for acoustic scattering from an impedance polygon 
facet}. The Journal of the Acoustical Society of America, 126(1), EL14-EL21, 2009.

\bibitem{Gaunard} G. Gaunard, {\it Sonar cross sections of bodies partially insonified by finite sound 
beams}, IEEE Journal of Oceanic Engineering, Vol OE-10 3, 1985.


\bibitem{George-PlaneFacets} O. George \& R. Bahl, {\it Simulation of backscattering of high frequency sound from 
complex objects and sand sea-bottom}, IEEE Journal of Oceanic Engineering, 20(2), 119-130, 1995.

\bibitem{Pignier} Pignier, Nicolas J., Ciarán J. O'Reilly, and Susann Boij. {\it A Kirchhoff approximation-based 
numerical method to compute multiple acoustic scattering of a moving source.}  Applied Acoustics 96, 108-117,  2015.

\bibitem{Sammelman} G.S. Sammelman, {\it Propagation and scattering in very shallow water}, Proc. IEEE Oceans 
\textbf{1}, 337-344, 2001. 

\bibitem{Wendelboe} G. Wendelboe, F. Jacobsen, \& J.M. Bell. {\it A numerically accurate and 
robust expression for bistatic scattering from a plane triangular facet}. The Journal of the Acoustical 
Society of America, 119(2), 701-704, 2005.

\bibitem{Abawi-Kirchhoff} A.T. Abawi, {\it Kirchhoff scattering from non-penetrable targets modeled 
as an assembly of triangular facets}, The Journal of the Acoustical Society of America, 140(3), 1878-1886,
2016.

\bibitem{penia} G. J. Macaulay, H. Peña, S. M. Fässler, G. Pedersen, \& E. Ona,  {\it Accuracy of the 
Kirchhoff-approximation and Kirchhoff-ray-mode fish swimbladder acoustic scattering models.}, PloS one 8 (5), 2013.

\bibitem{footeFrancis} Foote, Kenneth G., and David TI Francis. {\it Comparing Kirchhoff-approximation and 
boundary-element models for computing gadoid target strengths.} The Journal of the Acoustical Society of America 111 
(4),  1644-1654, 2002.

\bibitem{Volker} A.W.F. Volker \& M.G. Ter Morshuizen, {\it  Semi-analytical computation of the acoustic 
response of arbitrarily shaped objects.}, Contributed paper, European Congress on Acoustics, Sevilla, 2012.

\bibitem{Carley-quad} M. Carley. {\it Quadrature for second-order triangles in the Boundary Element Method}. arXiv 
preprint arXiv:1302.6054, 2013.

\bibitem{quadgk} QuadGK {\it Numerical integration in Julia}, {\it https://github.com/JuliaMath/QuadGK.jl}, 2017.

\bibitem{julia} Bezanson, J., Edelman, A., Karpinski, S., \& Shah, V. B. (2017). {\it Julia: A fresh approach to 
numerical computing}. SIAM review, 59(1), 65-98.

\bibitem{ts_subs} Schneider, H. G., Berg, R., Gilroy, L., Karasalo, I., MacGillivray, I., Morshuizen, M. T., \& Volker, 
A. {\it Acoustic scattering by a submarine: Results from a benchmark target strength simulation workshop}. ICSV10, 
2475-2482, 2003.

\bibitem{improved_betssi} Nell, C. W., \& Gilroy, L. E. (2003). {\it An improved BASIS model for the BeTSSi submarine}. 
DRDC Atlantic TR, 199, 2003.

\end{thebibliography}
\end{document}